 \documentclass[aip,jmp,amsmath,amssymb,preprint,author-numerical]{revtex4-1}

\usepackage{amsmath}
\usepackage{graphicx}
\usepackage{dcolumn}
\usepackage{bm}
\usepackage{graphics}
\usepackage{epsfig}
\usepackage{subfigure}
\usepackage{color}

\newcommand\eq{\begin{equation}}
\newcommand\be{\begin{equation}}
\newcommand\eeq{\end{equation}}
\newcommand\ee{\end{equation}}
\newcommand\ar{\begin{eqnarray}}
\newcommand\ear{\end{eqnarray}}

\newcommand{\ii}{{\rm i}}

\begin{document}

\newcommand{\nm}{\mbox{ nm}}
\newcommand{\micron}{\mbox{ $\mu$m}}
\newcommand{\fm}{\mbox{ fm}}
\newcommand{\Hz}{\mbox{ Hz}}
\newcommand{\watt}{\mbox{ W}}
\newcommand{\m}{\mbox{ m}}
\newcommand{\cm}{\mbox{ cm}}
\newcommand{\km}{\mbox{ km}}
\newcommand{\seconds}{\mbox{ s}}
\newcommand{\minutes}{\mbox{ min}}
\newcommand{\volts}{\mbox{ V}}
\newcommand{\MeV}{\mbox{ MeV}}
\newcommand{\eV}{\mbox{ eV}}
\newcommand{\meV}{\mbox{ meV}}
\newcommand{\rad}{\mbox{ radians}}
\newcommand{\Kelvin}{\mbox{ K}}
\newcommand{\Siemen}{\mbox{ S}}
\newcommand{\angstrom}{\mbox{ $\AA$}}


\title{Active plasmonic switching at mid-infrared wavelengths with graphene ribbon arrays}

\author{Hong-Son Chu}
\email{chuhs@ihpc.a-star.edu.sg}
\affiliation { 
Electronics and Photonics Department, A*STAR's Institute of High Performance Computing, Singapore
}  

\author{Choon How Gan}%
\affiliation{ 
College of Engineering, Mathematics and Physical Sciences, University of Exeter, Exeter EX4 4QF, United Kingdom 
}%


\begin{abstract}
An active plasmonic switch based on single- and few-layer doped graphene ribbon array operating in the mid-infrared spectrum is investigated with theoretical and numerical calculations. It is shown that significant resonance wavelength shifts and modulation depths can be achieved with a slight variation of the doping concentration of the graphene ribbon. The few-layer graphene ribbon array device outperforms the single-layer one in terms of the achievable modulation depth. Our simulations reveal that, by modulating the Fermi-energy level between 0.2 eV and 0.25 eV, a four-layer graphene ribbon array device can achieve a modulation depth and resonance wavelength shift of $\sim$13 dB and 0.94 $\mu$m respectively, compared to $\sim$2.8~dB and 1.85~$\mu$m for a single-layer device. Additionally, simple fitting models to predict the modulation depth and the resonance wavelength shift are proposed. These prospects pave the way towards ultrafast active graphene-based plasmonic devices for infrared and THz applications.

%
\end{abstract}

\maketitle

Recently, there is tremendous scientific and technological interest in the mid-infrared spectral range of 2-20 $\mu${m}. This range provides many potential applications for optics/photonics such as spectroscopy, materials processing, chemical and biomolecular sensing, remote explosive detection and covert communication systems~\cite{edi, sor}. Thanks to the 
integration with electronic devices and possibility to design devices with active control over the surface plasmon resonance~\cite{raether} at metal/dielectric interfaces, the mid-infrared spectral region is also attractive for the study of plasmonic devices~\cite{stan, yu}. However, due to a relatively weak refractive index change with electrical-bias, mechanical force or temperature, the active plasmonic devices typically exhibit low optical performance such as high power consumption or slow response time. 
Graphene, a single layer of carbon atoms gathered in a honeycomb lattice, exhibits many unique physical features and has been recently investigated for nanoscale optoelectronic integrated circuits. 
Graphene-based plasmonic nanostructures support highly confined plasmonic modes that can be tuned via chemical or eletrostatic doping, and are promising to serve as future platforms for highly integrated active plasmonic devices ranged from the infrared to THz frequencies~\cite{gan1,grig,ju,peresepl92,gan2,bao,yan,liu&cai,li&yu}. Therefore, it is important to develop broadly tunable plasmonic devices, either as an enabling technology or to add functionality to current plasmonic technologies. 
In this paper, we demonstrate with numerical simulations that an effective active plasmonic switch operating in the mid-infrared wavelength range can be realized with a doped graphene ribbon array patterened on a silicon-on-insulator substrate. We show that an increase of the Fermi energy level $E_F$  from 0.2 eV to 0.25 eV (i.e., increasing the carrier concentration from $2.9 \times 10^{12}$ to $4.6 \times 10^{12}$ cm$^{-2}$) can result in a significant shift of the surface plasmon resonance wavelength ($\Delta\lambda$). 
Moreover the modulation depth ($MD$) is significantly enhanced in the case of few-layer graphene ribbon arrays (number of graphene layers $N \lesssim 6$)~\cite{yan,ferrarinl}. In particular, the 
four-layer graphene ribbon array ($N = 4$) exhibits a modulation depth of 13 dB, which is $\sim$10 times better than that obtained with the single-layer graphene ribbon. Simple fitting models that accurately predict the modulation depth and the resonance wavelength shift as a function of the number of graphene layers $N$ are also proposed. It is worth noting that the few-layer structure allows the resonance wavelength to be blue-shifted to the near-infrared region,  opening the possibility to implement the device in that spectral regime.

\begin{figure}[ht]
  \centering
  \epsfig{file=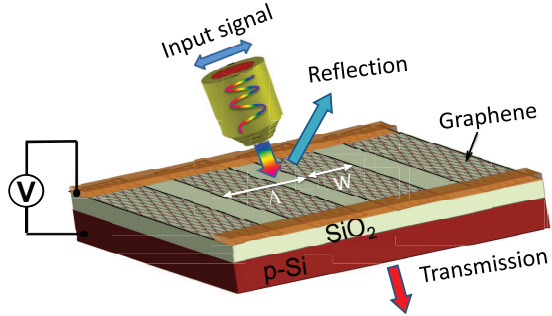, width=3.0 in}
  \caption{Schematic illustration of the electro-optical graphene ribbon-based plasmonic switch in mid-infrared wavelengths. The graphene ribbon array has the width of $w$ and period of $\Lambda$ on top of the SiO$_2$/p-Si substrate. }\label{Fig1}
  \end{figure}

The schematic illustration of the proposed graphene ribbon array-based plasmonic switch is shown in Fig.~\ref{Fig1}. The period-to-width (${\Lambda}/{w}$) of the graphene ribbons is fixed to be 2 in this paper. The graphene ribbon is deposited on top of a silicon-on-insulator substrate, following Ref.~\cite{ju}. To allow for doping through electrostatic gating, either a p-type silicon film~\cite{grig} (considered in this paper) or an ITO film~\cite{fang} can be employed as a bottom-gate, with two metallic pads acting as drain and source. A normally incident light polarized perpendicular to the ribbons is used to excite the surface plasmon mode~\cite{ju,nik}.

At room temperature $T=300$~K and for  mid-infrared wavelengths, the conductivity of single-layer graphene may be approximated with a Drude-like expression~\cite{yan,gan2} 
\eq\label{sigma}
\sigma \approx  \frac {\ii e^2 E_F} {\pi \hbar^2 (\omega + \ii \tau^{-1})} \,,
\eeq 
where $\omega$ is the angular frequency, and the electron relaxation time $\tau$ is taken to be 0.3 ps based on typical values of the carrier mobility~\cite{stauberprb76}.

We first investigate the resonance wavelength of the single-layer graphene ribbon array as a function of the graphene ribbon width and Fermi-level. The resonance wavelength of the graphene ribbon array $\lambda_0$ can be derived from the quasi-static analysis~\cite{ju,nik} and expressed as
\eq\label{resfreq}
\lambda_0 \approx {\frac{2 \pi c \hbar}{e}} \sqrt{\frac{\eta \epsilon_{\rm eff} \epsilon_0 w}{E_F}} \,
\eeq 
where $\epsilon_{\rm eff}$ is the effective permittivity of the medium surrounding the graphene ribbon array and taken to be $\epsilon_{\rm eff} = (\epsilon_{\rm SiO_2} + 1)/2$. The dimensionless constant $\eta = 3.1$ is a fitting parameter deduced from numerical simulations.

\begin{figure}[ht]
  \centering
  \epsfig{file=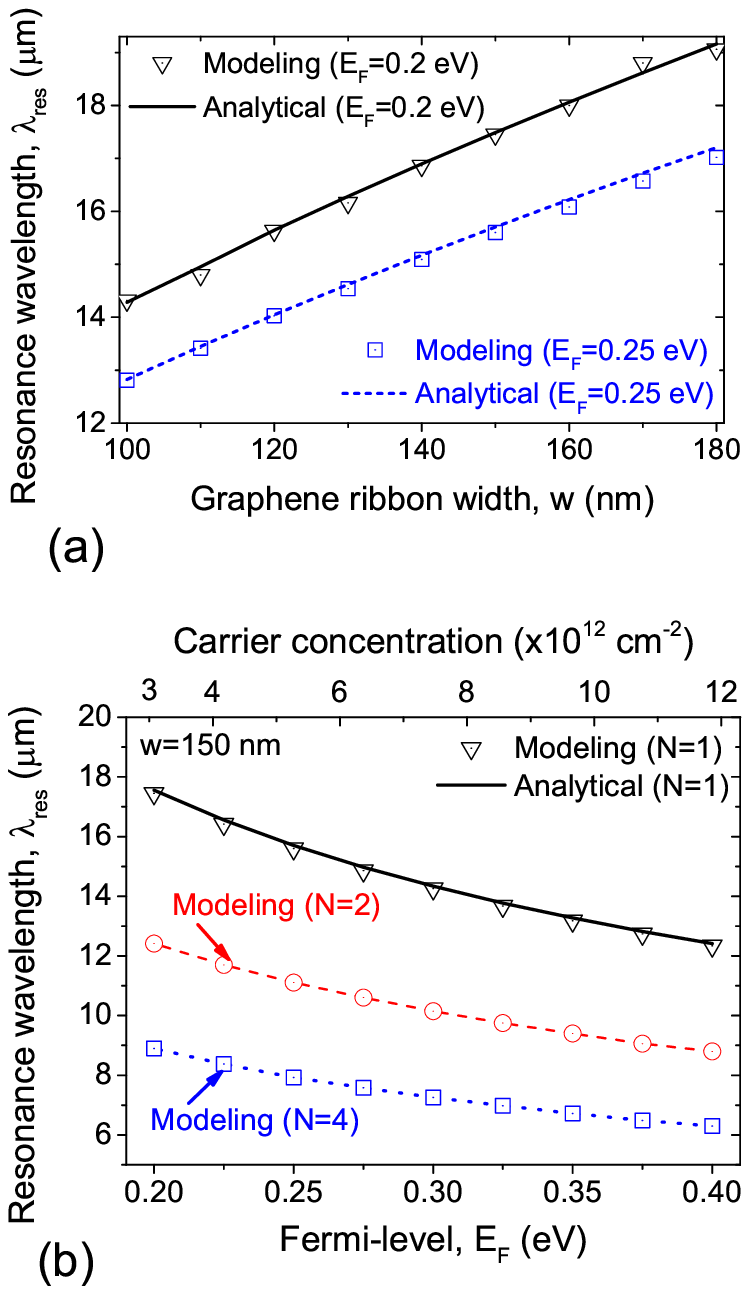, width=3.0 in}
\caption{Variation of the resonance wavelength with (a) graphene ribbon width $w$ for $E_{F}$ = 0.2 eV and 0.25 eV, and (b) Fermi-level $E_{F}$ for $N = 1, 2,$ and 4 with $w$ taken to be 150 nm. Solid lines and triangles are results from Eq.~\eqref{resfreq} and FDTD simulations, respectively. The circle and square dotted lines represent the modeling result for $N$ = 2 and 4.}\label{Fig2}
  \end{figure}

In Fig.~\ref{Fig2}.a, 
we compare the resonance wavelength of the single-layer graphene ribbon array obtained from Finite-difference time-domain (FDTD) method and Eq.~\eqref{resfreq}
for different ribbon width $w$ and two values of the Fermi-level $E_F = 0.2$~eV and $E_F = 0.25$~eV. 
Similarly, in Fig.~\ref{Fig2}.b, the resonance wavelength as a function of the Fermi-level $E_F$, ranged from 0.2 eV to 0.4 eV, is calculated both using FDTD numerical modeling and Eq.~\eqref{resfreq} for the case $w$=150 nm. 
Both methods yield results in reasonably good agreement. The resonance wavelength is clearly blue-shifted for narrower ribbon widths or higher Fermi-levels~\cite{ju}.
Thus, the resonance wavelength of the graphene ribbon array can be controlled by: (i) passive tuning (changing the ribbon width $w$), and/or (ii) active tuning (in-situ variation of the Fermi-level or carrier concentration).

As several works have been proposed to practically realize the doping of few-layer graphene including the chemical- and electrical-induced doping level of graphene~\cite{kwon,gune,liu,li,ryz}, we have also investigated the optical performance of the proposed device with $N > 1$.  
Here, it is taken that the conductivity for $N-$layer graphene is $N \sigma$~\cite{yan,ferrarinl}. 
In Fig.~\ref{Fig2}.b, the resonance wavelength as 
a function of the Fermi-level ranged from 0.2 to 0.4 eV for single-layer ($N = 1$), two-layer ($N = 2$) and four-layer graphene ribbon ($N = 4$) is shown. 
In shifting the Fermi-level from 0.2 eV to 0.25 eV,
the resonance wavelength shift  $\Delta \lambda$ is approximately $\sim$0.94, 1.3 and 1.85 $\mu$m for $N = 4, 2,$ and 1, respectively. 
The few-layer graphene structure exhibits a similar trend as the single-layer structure in terms of the variation of the resonance wavelength with $w$ and $E_F$.
Additionally, it is seen that increasing the number of graphene layers can further decrease the resonance wavelength towards the shorter wavelength spectrum region. This makes highly doped few-layer graphene ribbon structure a potential platform for integrated silicon photonic devices and for near-IR data communications applications.

\begin{figure}[ht]
  \centering
  \epsfig{file=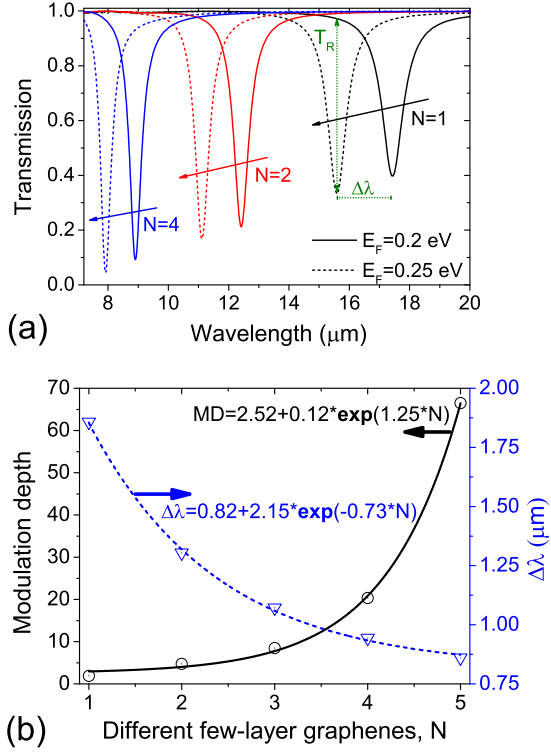, width=3.0 in}
  \caption{Performance of the proposed device at mid-infrared wavelengths for a shift in the Fermi-level from 0.2 eV to 0.25 eV. (a) Transmission for single-layer ($N = 1$), two-layer ($N = 2$) and four-layer graphene ribbon array ($N = 4$) as a function of wavelength. It is observed that the increase in the Fermi-level blueshifts the resonance wavelength, and that the modulation depth increases with the number of graphene layers ($N$). (b) Modulation depth and the resonance wavelength shift as a function of $N$.}\label{Fig3}
  \end{figure}

Next, let us look at the effect of the few-layer graphene ribbon structures on the optical performance of the proposed active plasmonic switch and modulator. In Fig.~\ref{Fig3}.a the simulated transmission for three sets of 
graphene ribbon array ($N = 1, 2,$ and 4) modulated between $E_F$=0.2 and 0.25 eV is plotted for the mid-infrared wavelengths. Comparing to the single-layer graphene ribbon array, the few-layer graphene exhibits a sharper resonance, leading to a larger modulation depth ($MD$). It is noted that $MD$ is defined as~\cite{bar,cai} 
\eq\label{MDdef}
MD = \bigg| \frac {T_{\rm on}-T_{\rm off}}{T_{\rm on}} \bigg|=\bigg|{1-T_{\rm R}}\bigg|, \,
\eeq 
where $T_{\rm on}$ and $T_{\rm off}$ are the transmission magnitude at the ``on'' state ($E_F$=0.25 eV) and ``off'' state ($E_F$=0.2 eV), respectively. 
From Fig.~\ref{Fig3}.a, the $MD$ at the resonance are $\sim$20.3 (13 dB), $\sim$4.8 (6.8 dB) and $\sim$1.92 (2.8 dB) for $N = 4, 2,$ and 1, respectively. From calculations (not shown in the figure), the $MD$ is even better when modulating with higher Fermi-levels. For instance, taking the ``on'' state as $E_F$=0.35 eV and ``off'' state as $E_F$=0.3 eV, the $MD$ obtained was $\sim$13.5, 9 and 4.4 dB for $N = 4, 2,$ and 1, respectively. However, as indicated by Eq.~\ref{resfreq}, the resonance wavelength shift is proportional to ${E_F}^{-1/2}$ and will be decreased when modulating with higher Fermi-levels.

We have numerically characterized the modulation depth and resonance wavelength shift as a function of $N$. The results for modulating between $E_F = 0.2$ eV to 0.25 eV are shown in Fig.~\ref{Fig3}.b. It is observed that increasing the number of graphene layers from 1 to 5 layers significantly increase the modulation depth approximately from $\sim$1.92 (2.8 dB) to $\sim$67.5 (18.3 dB). Evidently, if a higher Fermi-level or a larger carrier concentration is used, the $MD$ will be much higher. To establish a design guideline for the few-layer graphene structure, we have employed an exponential curve fitting procedure to develop simple empirical expressions for $MD$ and $\Delta{\lambda}$ as a function of $N~(N \lesssim 6)$:
\begin{subequations}\label{expfit}
\begin{align}
MD_{\rm fit} & \approx 3.52+0.12{\textbf{exp}}(1.25N)\,, \label{MDfit} \\
\Delta{\lambda}_{\rm fit} & \approx 0.82+2.15{\textbf{exp}}(-0.73N)\,. \label{resfit}
\end{align}
\end{subequations} 
It is seen from Eq.~\eqref{MDfit} that $MD$ increases exponentially with the number of graphene ribbon layers. This finding is indeed consistent with the results reported in Refs.~\cite{yan} and~\cite{thon}. Namely  in the mid-infrared and THz wavelengths, complete light absorption, i.e. near-zero transmission of light, can be achieved with either more layers of doped graphene or with periodically patterned graphene structures.

In conclusion, we have demonstrated that the use of single- and few-layer graphene ribbon array to effectively design a plasmonic switch in mid-infrared wavelengths. The proposed few-layer graphene ribbon dramatically outperforms the single-layer one in terms of the modulation depth. Fitting models that predict the modulation depth and the resonance wavelength shift to act as design guidelines are proposed and discussed. The proposed few-layer graphene ribbon structure could be useful for research in compact and largely tunable  mid-infrared photonic devices to realize on-chip CMOS optoelectronic systems.

\subsection*{} This work is supported by the National Research Foundation Singapore under its Competitive Research Programme (CRP Award No. NRF-CRP 8-2011-07).

\end{document}